\documentclass[aps,prb,twocolumn,superscriptaddress]{revtex4}

\input epsf.sty

\begin{document}

%
%

\title{Resonant inelastic x-ray scattering study of overdoped La$_{2-
x}$Sr$_{x}$CuO$_{4}$}

\author{S. Wakimoto}
\email[Corresponding author: ]{swakimoto@neutrons.tokai.jaeri.go.jp}
\affiliation{ Department of Physics, University of Toronto, Toronto,
   Ontario, Canada M5S~1A7 }
\affiliation{Advanced Science Research Center, 
   Japan Atomic Energy Research Institute, Tokai, Ibaraki 
   319-1195, Japan }

\author{Young-June Kim}
\affiliation{ Department of Physics, University of Toronto, Toronto,
   Ontario, Canada M5S~1A7 }
\affiliation{ Department of Physics, Brookhaven National Laboratory,
   Upton, New York 11973, USA }

\author{Hyunkyung Kim}
\affiliation{ Department of Physics, University of Toronto, Toronto,
   Ontario, Canada M5S~1A7 }

\author{H. Zhang}
\affiliation{ Department of Physics, University of Toronto, Toronto,
   Ontario, Canada M5S~1A7 }

\author{T. Gog}
\affiliation{ CMC-CAT, Advanced Photon Source, Argonne National
   Laboratory, Argonne, Illinois 60439, USA }

\author{R. J. Birgeneau}
\affiliation{ Department of Physics, University of Toronto, Toronto,
   Ontario, Canada M5S~1A7 }
\affiliation{ Department of Physics, University of California,
   Berkeley, Berkeley, California 94720-7300, USA }

\date{\today}

\begin{abstract}

Resonant inelastic x-ray scattering (RIXS) at the copper K
absorption edge has been performed for heavily overdoped samples
of La$_{2-x}$Sr$_{x}$CuO$_{4}$ with $x= 0.25$ and $0.30$.  We 
have observed the charge
transfer and molecular-orbital excitations which exhibit 
resonances at incident energies of $E_i= 8.992$ and $8.998$~keV,
respectively.
From a comparison with previous results on undoped and
optimally-doped samples, we determine that the charge-transfer 
excitation energy increases monotonically as doping increases.  
In addition, the $E_i$-dependences of the RIXS spectral weight 
and absorption spectrum exhibit no clear peak at 
$E_i = 8.998$~keV in contrast to results in the underdoped 
samples.
The low-energy ($\leq 3$~eV) continuum excitation intensity 
has been studied utilizing the high energy resolution of 
0.13~eV (FWHM).  A comparison of the RIXS profiles at $(\pi ~0)$ 
and $(\pi ~\pi)$ indicates that the continuum intensity exists 
even at $(\pi ~\pi)$ in the overdoped samples, whereas it has 
been reported only at $(0 ~0)$ and $(\pi ~0)$ for the $x=0.17$ 
sample. Furthermore, we also found an additional excitation on 
top of the continuum intensity at the $(\pi ~\pi)$ and 
$(\pi ~0)$ positions.

\end{abstract}


\maketitle

\section{Introduction}

The essential physics of the hole-doped high-$T_c$ cuprates 
can be described as that of a doped Mott insulator, which 
exhibits several phases varying from antiferromagnetic insulator 
to metallic superconductor.  The elucidation of the basic 
physics of these various complicated phases by studying the 
elementary excitations could yield information essential 
for uncovering the mechanism of high-$T_c$ superconductivity.
Experimental techniques that can probe the charge excitations 
have been dramatically developed in recent years. In particular,
resonant inelastic x-ray scattering (RIXS) in the hard x-ray 
regime has attracted much attention due to its ability to probe 
the momentum dependence of the charge excitations. This technique 
was initially applied to NiO;~\cite{Kao_96} in that case a large 
enhancement of the inelastic signal with incident photon energy 
was observed near the Ni K edge.  More recently this technique 
has been applied to various cuprates including high-$T_c$
materials~\cite{Hill_98,Abbamonte_99,
Hasan_sci00,Kim_02,Kim_04_CT,Kim_04_MO,Hasan_04,Ishii_ybco,
Ishii_ncco,Lu_05} and other highly-correlated electron
systems.~\cite{Hasan_02, Kim_04_LiCuO,Kim_04_SrCuO}

A recent RIXS study near the Cu K edge in undoped La$_2$CuO$_4$ 
(LCO) has revealed three features in its charge excitation 
spectrum.~\cite{Kim_02} Two features, labeled A and B, appear at
energy transfers $\omega = 2.2$~eV and 3.9~eV, respectively; 
these features show the largest intensity enhancement at an 
incident photon energy of $E_i = 8.991$~keV. The other feature, 
labeled C, appears at $\omega = 7.2$~eV, and resonates at the 
higher incident energy of $E_i = 8.998$~keV.  Since the resonance 
energy of the A and B features corresponds to the absorption 
feature associated with a well-screened state,~\cite{Li_91} these 
two have been attributed to the charge-transfer (CT) excitations 
from the O $2p$ to the Cu $3d$ upper Hubbard band.  On the other 
hand, the resonance energy of feature C is associated with a 
poorly screened state, and has been attributed to a molecular 
orbital (MO) excitation that involves a transition between
bonding and anti-bonding states of the Cu $3d$ and O $2p$
orbitals.~\cite{Kim_04_MO}

Recently, there has been a number of RIXS studies of doped cuprate
superconductors.~\cite{Kim_04_CT,Kim_04_MO,Hasan_04,Lu_05} 
For the La$_{2-x}$Sr$_{x}$CuO$_{4}$ (LSCO) system, Kim 
{\it et al.}~\cite{Kim_04_CT} showed that the two features of 
the CT excitations in LCO are
also observed for $x=0.05$, but only a single feature is
observed at $\sim 4$~eV for $x=0.17$. In addition, a
continuum-like excitation appears for $x=0.17$ below the charge
transfer excitation energy ($\omega < 2$~eV) at the zone center (0
~0) and at the zone boundary $(\pi ~0)$.~\cite{Kim_04_CT}  The zone
center result is consistent with optical measurements which
show a transfer of the spectral weight from the CT excitation to
lower energies.~\cite{Uchida_91}   For the MO excitation, the
excitation energy increases with doping.~\cite{Kim_04_MO}  This was
attributed to the decreased Cu-O distance in doped samples, which
leads to an increased energy splitting between the bonding and
antibonding molecular-orbitals due to the increased $p-d$
hybridization.

Although measurements have been carried out for underdoped and 
optimally doped samples, the overdoped region has not yet been 
studied with RIXS. Such studies should yield important insights 
into the nature of the charge excitations observed with the RIXS
technique. We note that thus far LSCO appears to be the only 
high-$T_c$ system where large, high quality heavily overdoped 
crystals are available.
In this paper, we report RIXS results for overdoped LSCO with 
$x=0.25$ and 0.30; the primary purpose of this study is to 
elucidate the evolution of the charge excitations in the overdoped 
region.  We find that the CT and MO excitations in the overdoped 
compounds exhibit resonantly enhanced intensity over a broader 
range of incident energies than those of an optimally-doped and 
undoped (parent) compound.  We also observe that the near-edge 
x-ray absorption spectrum exhibits a feature associated with 
the well-screened final state, which is well-defined similar to
its counterpart in underdoped samples. On the other hand, we do not
find a well-defined peak corresponding to the poorly-screened final
state. Profiles measured utilizing a high energy resolution 
spectrometer configuration indicate 
that the continuum excitation exists below 3~eV at $(\pi ~\pi)$ in
addition to the $(0 ~0)$ and $(\pi ~0)$ positions. 
On top of the continuum intensity, we observed an additional 
excitation at $\omega \sim 1.8$~eV, which exhibits resonantly 
enhanced intensity around $E_i \sim 8.993$~keV.

The paper is organized as follows.  Section II describes the
experimental details.  The CT and MO excitations measured 
with a conventional RIXS setup are presented Sec. III, while a 
detailed study of the continuum intensity below 3~eV using the 
high resolution setup is reported in Sec. IV.  Finally the results 
are summarized in Sec. V.  Since the results for the $x=0.25$ and 
$0.30$ crystals are quantitatively similar, we do not distinguish
between these two samples in our discussion unless there is a
special need.

\section{Experimental details}

Single crystals of LSCO with $x=0.25$ and
$0.30$ for the RIXS measurements were grown by the 
traveling-solvent floating-zone (TSFZ) method in
the same manner as those used for neutron-scattering
experiments.~\cite{waki_04PRL,waki_full}  The grown crystal rods 
have a typical size of 6~mm in diameter and 80~mm in length; they 
were cut into pieces 5~mm $\times$ 4~mm $\times$ 1~mm along the 
tetragonal $a$, $c$ and $b$-directions, respectively, and annealed 
at 850~$^{\circ}$C for 12 hours in flowing oxygen to compensate 
for oxygen deficiencies.  The temperature dependence of the 
superconducting shielding fraction determined by magnetization 
measurements shows that the $x=0.25$
sample has $T_c = 15$~K whereas the $x=0.30$ sample shows no bulk
superconductivity down to 2~K. The $ac$ surfaces, which the
x-ray beams in the RIXS experiments are incident upon, were 
polished using a 1 $\mu$m grit size.

The RIXS experiments were performed at the undulator beam line 
9IDB at the Advanced Photon Source at Argonne National Laboratory.
For the measurements of the CT and MO excitations, we used the Si 
(333) reflection from the double bounce main monochromator, together 
with a spherical Ge(733) analyzer with 1~m radius
of curvature. This setup gives an instrumental energy resolution of
0.4~eV full-width at half maximum (FWHM). For the measurement
of the continuum intensity below $3$~eV, a Si(444) channel-cut
secondary monochromator was used in addition to a Si(111) primary
monochromator. A spherical Ge(733) analyzer with a 2~m radius of
curvature was also utilized.  This spectrometer configuration yielded 
a net energy resolution of
0.13~eV (FWHM).  More technical details of the finer resolution 
setup has been reported in Ref.~\onlinecite{Hill_05}.

The energy resolution has been measured by scanning an incoherent 
background signal along the energy direction.  For the finer 
resolution setup, we cooled the samples down to 9~K by a closed 
cycle refrigerator to reduce a possible phonon contribution to 
the elastic background intensity.  However we found that the elastic 
peak width along the energy direction does not show clear temperature 
dependence.  Thus, we conclude no phonon effect on the elastic 
peak.

For all measurements, the scattering plane was vertical and the 
polarization of the incident x-ray was fixed along the $c$-axis.  
Typical scans were performed by counting for 2 minutes at each data 
point, and the counts were normalized to monitor counts 
corresponding to 1 s.
In our analysis, we have used the following background subtraction
and normalization method. The largest source of background arises
from the contribution of the elastic ($\omega=0$) intensity. The 
energy gain ($\omega < 0$) side spectrum is expected to represent 
purely background intensity, since the detailed balance factor at 
this temperature and energy scale vanishes. Therefore, the background
subtracted intensity can be written as $D(\omega) = I(\omega) -
I(-\omega)$ $(\omega > 0)$, where $I(\omega)$ is the raw intensity,
assuming that the elastic background is symmetric in $\omega$.
Unless noted otherwise, we used the intensity of the Cu K$\beta_5$
emission line as a normalization factor, since this emission
intensity is expected to be proportional to the sample volume
probed.

At last, throughout this paper,
we use tetragonal notation with $a = 3.77$~\AA~ along the Cu-O-Cu
direction, corresponding to $a^* = 1.67$~\AA$^{-1}$.

\section{Charge transfer and molecular-orbital excitations}

First, we present the experimental results for the main spectral
features, which correspond to the CT and the MO excitations. Figure
1 shows the evolution of the RIXS spectra as a function of the incident
energy $E_i$ near the Cu K edge at the positions ${\rm \bf Q} = 
(3.05, 0)$ corresponding to a position near the zone center 
${\rm \bf q} = (0.1\pi ~0)$ and ${\rm \bf Q} = (2.5, 0)$ corresponding 
to the zone boundary position ${\rm \bf q} = (\pi ~0)$. Unfortunately, 
the zone center $(3, 0, 0)$ position is contaminated at $\omega = 0$ 
presumably by the tails of the Bragg peaks at (3, 0, 1) and 
(3, 0, $\bar{1}$), so that the data have been obtained at positions 
slightly away from the zone center. However, a relatively high elastic 
intensity remains at $(0.1\pi ~0)$, producing a broader tail than that 
at $(\pi ~0)$. Nevertheless, a similar incident energy dependence of 
the RIXS spectra is observed at both positions with a CT excitation 
around 4~eV and the MO excitation around 8~eV, exhibiting resonance 
enhancements at $E_i \sim 8.992$~keV and $\sim 8.998$~keV, respectively.

\begin{figure*}
\centerline{\epsfxsize=5.5in\epsfbox{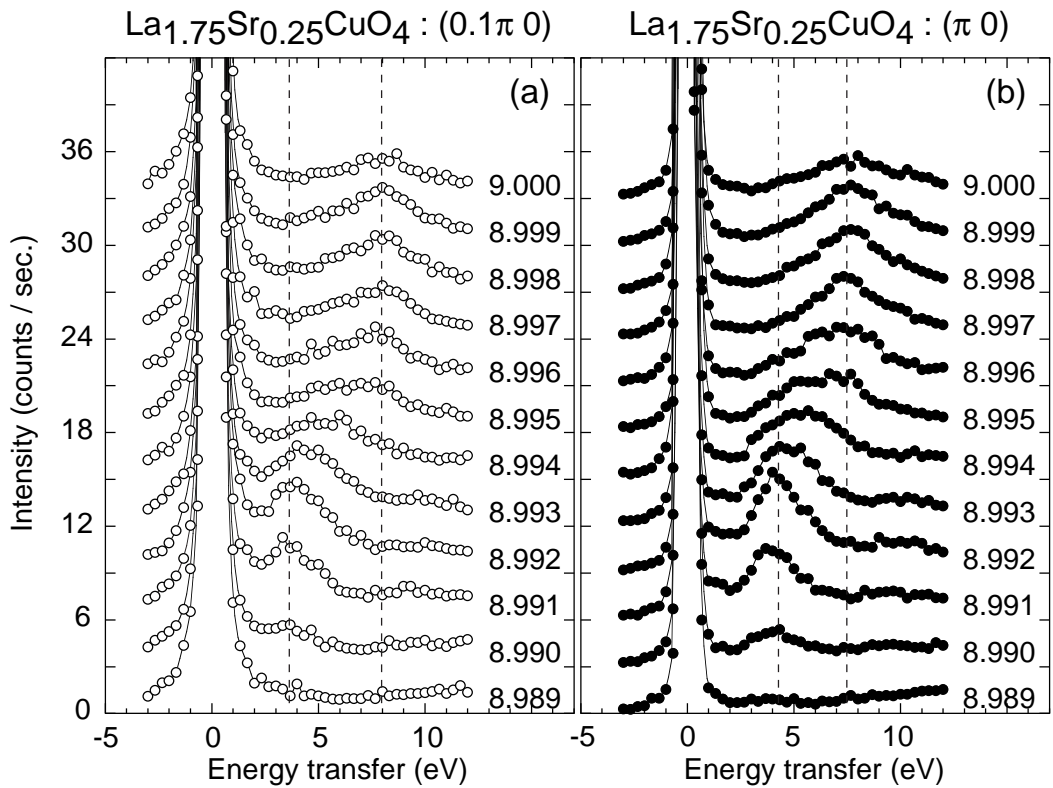}}
\caption{$E_i$-variation of the room-temperature RIXS spectra of the
$x=0.25$ sample at (a) a position near the zone-center $(0.1\pi ~0)$ and (b)
zone-boundary $(\pi ~0)$.  These are measured at ${\rm Q}=(3.05, 0,
0)$ and $(2.5, 0.5, 0)$, respectively.  The data are shifted vertically
for clarity.  Dashed lines represent energies of the CT and the MO
excitations.}
\end{figure*}

To study any possible change of the RIXS spectra as a function of
momentum and doping, the spectral weight has been calculated by
simply summing up the background subtracted and normalized intensity
between 2~eV and 12~eV. Spectral weights so obtained for the data at
$(0.1\pi ~0)$ and $(\pi ~0)$ are shown in Fig.~2(a) by circles and
squares, respectively. Although the (near) zone center data show smaller
spectral weights, the $E_i$ dependence appears to be very similar at
both positions. This is more clearly seen in Fig.~2(b) which shows 
the $E_i$ dependences of the spectral weights normalized to their
maximum values. Thus, although the excitation energies are slightly
different at different ${\rm \bf q}$ positions as shown in Fig.~1,
the resonance behavior is similar.

\begin{figure}
\centerline{\epsfxsize=3.3in\epsfbox{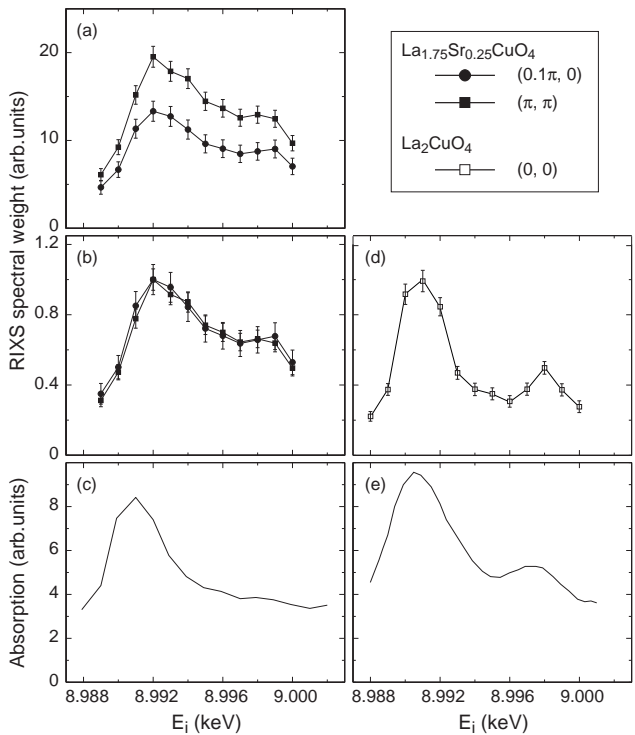}}
\caption{(a)
RIXS spectral weights for the $x=0.25$ sample as a function of
$E_i$.  The spectral weights are normalized to the effective volumes
estimated from the Cu-K$\beta_5$ intensity.  (b) RIXS spectral weights
normalized to the maximum value.  For figures (a) and (b), the squares
and circles represent data at $(0.1\pi ~0)$ and $(\pi ~0)$,
respectively.  (d) shows spectral weight of the parent material
La$_2$CuO$_4$ calculated from the data reported in Fig. 1 (a) of 
Ref.~\onlinecite{Kim_02}.  The data are normalized to the maximum value.
(c) and (e) indicate the absorption of $x=0.25$ and pure samples,
respectively, measured by monitoring the Cu-K$\beta_5$ intensity.
The data of (e) is referred from Fig. 1 (b) of Ref.~\onlinecite{Kim_02}.
Error bars in figures (a), (b) and (d) result from a summation of the
statistical errors of summed data points.}
\end{figure}

We observe that the spectral weight does not show a well-defined peak
at $E_i \sim 8.998$~keV, which is the MO excitation resonance
energy, whereas the spectral weight has a peak at $E_i \sim
8.992$~keV, the CT excitation resonance energy. We also found that
the x-ray absorption spectrum shown in Fig.~2(c), which is measured
by monitoring the intensity of Cu K$\beta_5$ line at
$E_f=8.973$~keV, does not exhibit a well-defined peak associated with
the so-called poorly-screened state around $8.998$~keV.
This result should be contrasted with the results in the undoped
samples. Figure~2(d) shows the spectral weight of La$_2$CuO$_4$
calculated in the same manner using the data reported in
Ref.~\onlinecite{Kim_02}.  It is clearly seen that the spectral
weight of La$_2$CuO$_4$ has two well-defined peaks at the energies
where the CT and the MO excitations resonate.  Consistent with this,
the absorption spectrum of La$_2$CuO$_4$ also has the well-defined peaks
for both well-screened and poorly screened states as shown in
Fig.~2(e).~\cite{Kim_02} Since such a two-peak structure has also been
observed for the $x=0.17$ sample,~\cite{Kim_04_CT} the
disappearance of the higher energy peak in the absorption spectrum
and the broad nature of the RIXS spectral weight distribution seem
to be characteristic of overdoped metallic samples.

\begin{figure}
\centerline{\epsfxsize=3.3in\epsfbox{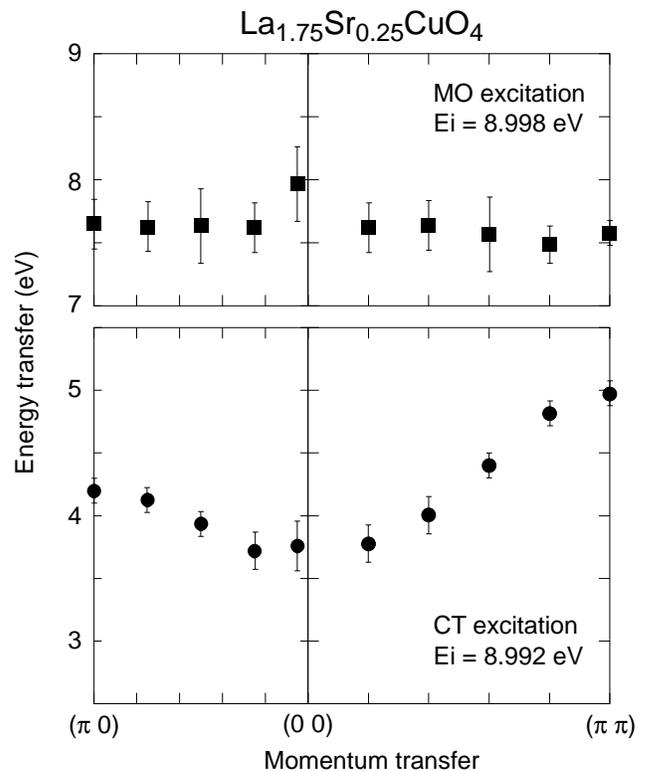}}
\caption{Dispersion relation of the MO excitation (upper) measured
with $E_i = 8.998$~keV and the CT excitation (lower) measured with
$E_i = 8.992$~keV for $x=0.25$.  Error bars are estimated from 
a simple fit to a Lorentzian function.}
\end{figure}

The dispersions of the CT and the MO excitations have been studied by
collecting the RIXS spectra at fixed $E_i=8.992$~keV and 8.998~keV
at several $q$-positions between $(3, 0)$ and $(2.5, 0)$, and
between $(3, 0)$ and $(2.5, 0.5)$.  The data have been fitted by a
Lorentzian function with a sloping background to determine the
excitation energies.  The dispersions so obtained are shown in
Fig.~3.  The CT excitation appears at 3.7~eV at the zone center and
reaches 4.2~eV at $(\pi ~0)$ and 5~eV at $(\pi ~\pi)$.  This
dispersion is qualitatively similar to that observed in the $x=0.17$
sample.~\cite{Kim_04_CT}  The actual profiles, however, show systematic
changes with increasing hole concentration.
The left panel of Fig. 4 indicates profiles of the CT peak for $x = 0$,
$0.17$ (from Refs.~\onlinecite{Kim_02} and ~\onlinecite{Kim_04_CT})
and $0.30$.  To compare the CT peak itself, the intensity is
normalized to the CT peak intensity. It may be clearly seen that the CT
peak shifts its weight to higher energy with increasing doping at
all $q$ positions.  This is consistent with an increase of the CT gap
with doping predicted by Tsutsui {\it et al.}~\cite{Tsutsui_03} and
agrees quantitatively with their numerical
calculations.~\cite{Tsutsui_04,Tsutsui_private}
The large value of this shift ($\sim 1$~eV) is somewhat surprising,
given that the chemical potential shift due to the hole doping observed 
in photoemission experiments is less than 0.5~eV.\cite{Ino_97} However, 
the upper Hubbard band shifts to higher energy as holes are
doped.~\cite{Tsutsui_private} This combined effect of the chemical
potential shift and the shift of the unoccupied band seems to 
contribute to the large overall shift of the CT feature.

It is also interesting to note that the large shift of the CT peak
energy in the overdoped regime may imply a large overlap of the O $2p$ 
and the lower Hubbard Cu $3d$ bands, resulting in the existence of 
partially doped-holes with Cu $3d$ character in the overdoped regime.  
This in turn would suppress magnetic correlations as 
reported by recent neutron scattering measurements of the overdoped 
samples.~\cite{waki_04PRL,waki_full}  
However, detailed numerical calculations are necessary 
to draw any firm conclusions.

\begin{figure}
\centerline{\epsfxsize=3.3in\epsfbox{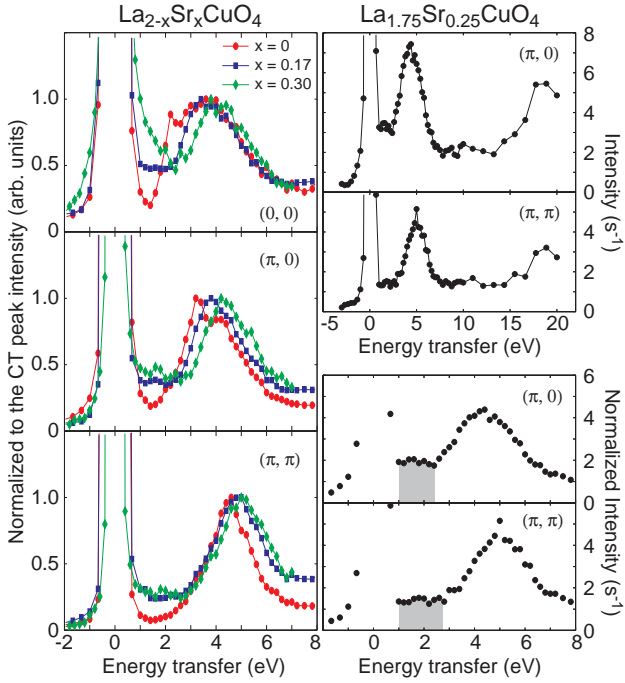}}
\caption{(Color
online) The left panel shows a comparison of the RIXS spectra for $x=0$,
0.17, and 0.30 at three different $q$-positions.  The data for $x=0$ 
and 0.17 are referred from Ref.~\onlinecite{Kim_02} and
\onlinecite{Kim_04_CT}.  The profiles are normalized to the CT peak
intensity.  The data for $x=0.30$ were taken utilizing the fine energy
resolution setup.  The right panel shows the RIXS spectra for $x=0.25$ 
with data as-measured (top) and normalized to the Cu K$\beta_5$ 
intensity (bottom).}
\end{figure}

The nature of the MO excitation also appears to differ from that
in the $x=0.17$ sample.~\cite{Kim_04_MO} Although the zone boundary
has a lower excitation energy than the zone center as is the case
for $x=0.17$, the dispersion of the overdoped sample is flat in
the large $q$ region as shown in Fig. 3.  As discussed in
Ref.~\onlinecite{Kim_04_MO}, the energy and the dispersion of the MO
excitation increases as the Cu-O distance decreases, and concomitantly 
the hybridization between the Cu $3d_{x^2-y^2}$ and the O $2p_{\sigma}$
orbitals becomes stronger. The excitation energy of 7.6~eV at $(\pi
~0)$ is very close to that of the $x=0.17$ sample which has the same
Cu-O distance as the $x=0.25$ sample, and, moreover, if we define the
dispersion as the difference between the energies at $(\pi ~0)$ and $(0
~0)$, the overall dispersion of $~\sim 0.4$~eV of the present sample is 
similar to that of the $x=0.17$ sample ($~\sim 0.5$~eV).

\section{Low-energy excitation}

The plots on the left side of Fig. 4 show clear evidence for the
appearance of a low energy excitation below the CT peak for the doped
samples. Although the excitation spectrum around 1~eV for the undoped
sample decreases to the background level (intensity at $\omega=-1$~eV),
the intensity at 1~eV remains well above the background in the
doped samples. As reviewed in Sec. 1, Kim {\it et
al.}~\cite{Kim_04_CT} have observed that the continuum intensity at
low energy ($1\sim 2$~eV) exists at (0 ~0) and at $(\pi ~0)$, but
not at $(\pi ~\pi)$. To test this in the overdoped sample, we have
compared the RIXS spectra for $x=0.25$ at $(\pi ~0)$ and $(\pi
~\pi)$.  The top-right panel of Fig. 4 shows the as-measured data of RIXS
spectra taken with $E_i = 8.992$~keV including the peak around 19~eV
due to the Cu K$\beta_5$ fluorescence emission. The bottom-right
panel shows the same spectrum after normalization using the Cu
K$\beta_5$ intensity. Both spectra exhibit a constant intensity between 
1 and 3 eV as shown by the shaded area, although the continuum intensity 
at the $(\pi ~\pi)$ position appears to be smaller than that at 
$(\pi ~0)$.  Thus, the data are suggestive of the existence of the 
continuum excitation even at $(\pi ~\pi)$ in the overdoped samples.

\begin{figure}
\centerline{\epsfxsize=3.3in\epsfbox{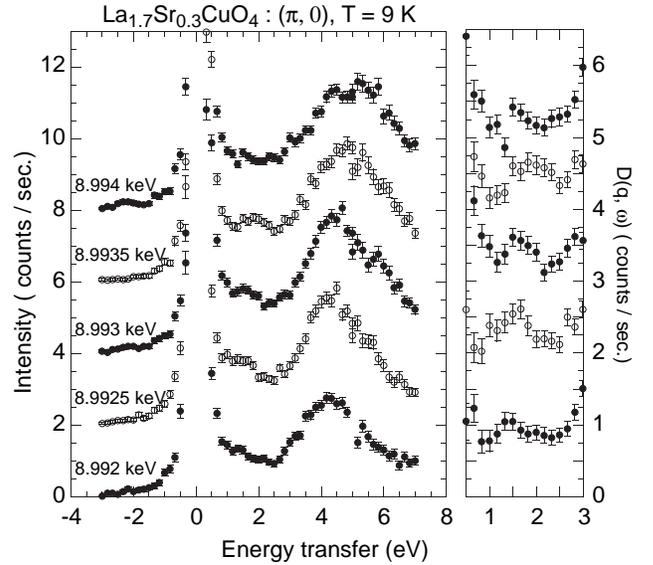}}
\caption{(Left)
RIXS spectra of the $x=0.30$ sample at $(\pi ~0)$ at various $E_i$ at 9 K.
Each spectrum is shifted by 2 counts for clarity. (Right) The 
differential intensity $D(q,~\omega)$ derived by subtracting the energy
gain side intensity as a background from the energy loss side. The data
are shifted by 1 count each.} \label{Fig:RIXS4}
\end{figure}

\begin{figure}
\centerline{\epsfxsize=3.3in\epsfbox{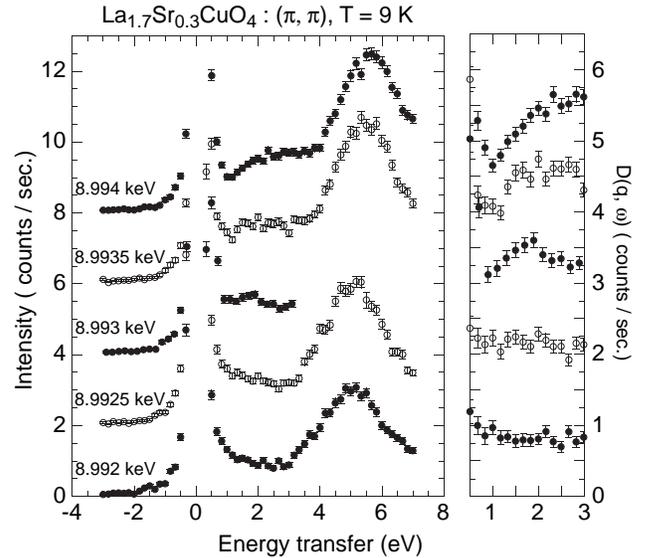}}
\caption{Analogous
plots to Fig.~\ref{Fig:RIXS4} for the RIXS spectra at $(\pi ~\pi)$.}
\end{figure}

We have also measured the RIXS spectra for $x=0.30$ at 9~K utilizing
the high resolution instrumentation ($\sim 0.13$~eV) which enables
us to study the detailed structure of the continuum excitation. Figure~5
shows the evolution of the RIXS spectra as a function of $E_i$ between
8.992 and 8.994~keV at $(\pi ~0)$. It appears that there is a small
peak on top of the continuum intensity at $\sim 1.5$~eV resonating in the
range of $8.9925 \leq E_i \leq 8.9935$~keV.  This may be more clearly
seen in the right panel of the figure, where the background
subtracted spectra are plotted on a different scale to emphasize the
continuum region. The data below 1~eV have a large ambiguity due to the
tail of the elastic peak, while the increase of $D(~\omega)$ above
2.5~eV is due to the CT peak. A peak at 1.5~eV is, nevertheless,
clear in the range of $8.992 \leq E_i \leq 8.993$~keV.

Similar features have been found at the $(\pi ~\pi)$ position.  
Figure 6 shows plots analogous to those in Fig.~5 for $(\pi ~\pi)$. 
In the right panel, the profile at $E_i = 8.993$~keV exhibits a peak 
of 0.5~counts/s appearing around $\omega = 1.8$~eV on top of the
continuum intensity of 1~count/s, whereas the data at $E_i =
8.992$ and $8.9925$~keV show only a flat continuum intensity.  As
$E_i$ increases to 8.994~keV, the peak appears to become broader and
larger, and seems to shift to higher energy.

We have studied the $q$-dependence of this excitation
around $(\pi ~\pi)$ with fixed $E_i$ at 8.993~eV.
Figure 7 shows the background-subtracted and normalized RIXS 
intensity $D(q, \omega)$ at $1.0 \leq \omega \leq 1.8$~eV
along the $(2.5+q, -0.5+q, 0)$ direction which is equivalent to the
trajectory in a reduced momentum space shown by an arrow in the
inset.  It is evident that a peak develops at $(\pi
~\pi)$ with increasing energy transfer and it reaches an intensity 
of $\sim 0.5$~counts/sec. at 1.8~eV on top of the $q$-independent 
continuum intensity of 1~count/sec, consistent with the data for 
$E_i = 8.993$~eV in Fig. 6.

It is open question what the origin of this excitation is.
From a consideration of the excitation energy of this feature, 
a possible origin is
the local $dd$-excitation which has been observed around 
1.5~eV.~\cite{Kuiper_98,Ghiringhelli_04}
However more detailed measurements are 
necessary to draw definitive conclusions.

\begin{figure}
\centerline{\epsfxsize=3.3in\epsfbox{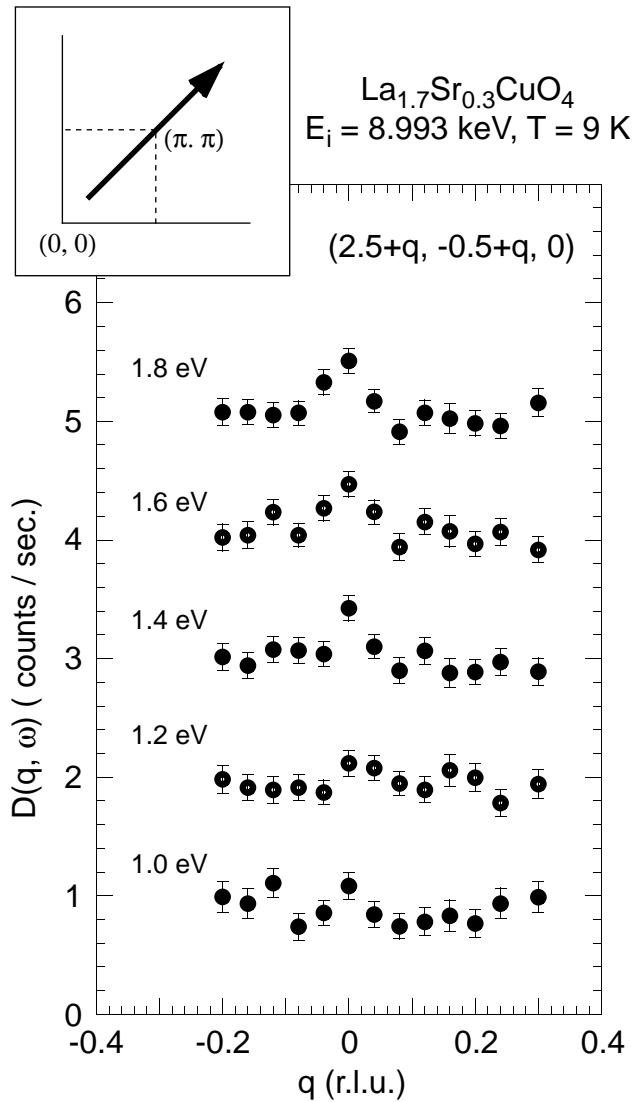}}
\caption{$q$-dependence of the RIXS intensity of the $x=0.30$ sample for 
$1 \leq \omega \leq 1.8$~eV.  The data are derived by normalizing to the
intensity at $\omega = 0$~eV after subtracting the energy gain side
intensity as a background.  Error bars result from a summation of the 
statistical errors of the energy loss signal and 
the energy gain background.   The data are shifted vertically for clarity.
The scanning direction is shown by an arrow in the inset.}
\end{figure}

\section{Concluding remarks}

We have studied heavily overdoped LSCO
with $x=0.25$ and 0.30 using the resonant inelastic x-ray scattering
technique near the Cu K edge. We observe several features of the
charge-transfer and molecular-orbital excitations in these
materials.
First, the resonance behavior in this overdoped sample appears to be
broader in incident photon energy than that of the undoped parent material.
Furthermore, in the absorption spectra of the overdoped samples, the 
peak corresponding to the poorly-screened state is damped.  
This presumably is related to the metallic nature of the overdoped
samples, and the better screening due to mobile charge carriers.
Second, we found that the CT peak energy increases with doping $x$.
This agrees with the prediction of an increase of the CT gap energy
with increasing $x$ and specifically, the numerical calculations of 
Tsutsui {\it et al.}~\cite{Tsutsui_03,Tsutsui_04} and Tsutsui.~\cite{Tsutsui_private}

Our study of the low energy excitations has shown that the continuum
intensity appears at all momentum transfers for the overdoped samples.
From detailed measurements utilizing the high energy resolution of
$0.13$~eV, 
we observed an additional excitation on the continuum intensity 
at $(\pi ~\pi)$ and $(\pi ~0)$,
which shows resonantly enhanced intensity at $E_i \sim 8.993$~keV.  
It will be important to study the dispersion of this feature, as well
as to determine whether or not this
excitation exists in the samples with lower doping to elucidate the origin
of this excitation.

\begin{acknowledgments}

The authors thank D. Casa, D. Ellis, J. P. Hill, K. Tsutsui, C. T. 
Venkataraman, and K. Yamada 
for invaluable discussions.  The work at the University of Toronto is part of the
Canadian Institute of Advanced Research and supported by Natural
Science and Engineering Research Council of Canada.  The work at BNL is
supported by the U. S. Department of Energy, Division of Material
Science, under contract No. DE-AC02-98CH10886.  The APS is
supported by the U. S. Department of Energy, Basic Energy Sciences,
Office of Science, under contract No. W-31-109-Eng-38.

\end{acknowledgments}


\begin{references}

\bibitem{Kao_96} C.-C. Kao, W. A. L. Caliebe, J. B. Hastings, and J.
M. Gillet, Phys. Rev. B {\bf 54}, 16361 (1996).

\bibitem{Hill_98} J. P. Hill, C.-C. Kao, W. A. L. Caliebe, M.
Matsubara,  A. Kotani, J. L. Peng, and R. L. Greene, Phys. Rev.
Lett. {\it 80}, 4967 (1998).

\bibitem{Abbamonte_99} P. Abbamonte, C. A. Burns, E. D. Isaacs, P. M.
Platzman, L. L. Miller, S. W. Cheong, and M. V. Klein, Phys. Rev.
Lett. {\bf 83}, 860 (1999).

\bibitem{Hasan_sci00} M. Z. Hasan, E. D. Isaacs, Z.-X. Shen, L. L.
Miller, K. Tsutsui, T. Tohyama, and S. Maekawa, Science {\bf 288},
1811 (2000).

\bibitem{Kim_02} Y. J. Kim, J. P. Hill, C. A. Burns, S. Wakimoto, R.
J. Birgeneau, D. Casa, T. Gog, and C. T. Venkataraman, Phys. Rev.
Lett. {\bf 89}, 177003 (2002).

\bibitem{Kim_04_CT} Y.-J. Kim, J. P. Hill, S. Komiya, Y. Ando, D.
Casa, T. Gog, and C. T. Venkataraman, Phys. Rev. B {\bf 70}, 094524
(2004).

\bibitem{Kim_04_MO} Y.-J. Kim, J. P. Hill, G. D. Gu, F. C. Chou, S.
Wakimoto, R. J. Birgeneau, S. Komiya, Y. Ando, N. Motoyama, K. M.
Kojima, S. Uchida, D. Casa, and T. Gog, 
Phys. Rev. B {\bf 70}, 205128 (2004).

\bibitem{Hasan_04} M. Z. Hasan, Y. Li, D. Qian, Y.-D. Chuang, H.
Eisaki, S. Uchida, Y. Koga, T. Sasagawa, and H. Takagi,
cond-mat/0406654.

\bibitem{Ishii_ybco} K. Ishii, K. Tsutsui, Y. Endoh, T. Tohyama,
K. Kuzushita, T. Inami, K. Ohwada, S. Maekawa, T. Masui, S. Tajima,
Y. Murakami, and J. Mizuki, Phys. Rev. Lett. {\bf 94}, 187002 (2005).

\bibitem{Ishii_ncco} K. Ishii, K. Tsutsui, Y. Endoh, T. Tohyama, 
S. Maekawa, M. Hoesch, K. Kuzushita, M. Tsubota, T. Inami, 
J. Mizuki, Y. Murakami, and K. Yamada,  Phys. Rev. Lett. {\bf 94}, 207003 (2005).

\bibitem{Lu_05} L. Lu, G. Chabot-Couture, X. Zhao, J. N. Hancock, 
N. Kaneko, O. P. Vajk, G. Yu, S. Grenier, Y. J. Kim, D. Casa, 
T. Gog, and M. Greven, Phys. Rev. Lett. {\bf 95}, 217003 (2005).

\bibitem{Hasan_02} M. Z. Hasan, P. A. Montano, E. D. Isaacs, Z.-X.
Shen, H. Eisaki, S. K. Sinha, Z. Islam, N. Motoyama, and S. Uchida,
Phys. Rev. Lett. {\bf 88}, 177403 (2002).

\bibitem{Kim_04_LiCuO} Y.-J. Kim, J. P. Hill, F. C. Chou, D. Casa, T.
Gog, and C. T. Venkataraman, Phys. Rev. B {\bf 69}, 155105 (2004).

\bibitem{Kim_04_SrCuO} Y.-J. Kim, J. P. Hill, H. Benthien, F. H. L.
Essler, E. Jeckelmann, H. S. Choi, T. W. Noh, N. Motoyama, K. M.
Kojima, S. Uchida, D. Casa, and T. Gog, Phys. Rev. Lett. {\bf 92},
137402 (2004).

\bibitem{Li_91} C. Li, M. Pompa, A. C. Castellano, S. D. Longa, and
A. Bianconi, Physica (Amsterdam) {\bf 175C}, 369 (1991).

\bibitem{Uchida_91} S. Uchida, T. Ido, H. Takagi, T. Arima, Y.
Tokura, and S. Tajima, Phys. Rev. B {\bf 43}, 7942 (1991).

\bibitem{waki_04PRL} S. Wakimoto, H. Zhang, K. Yamada, I. Swainson,
Hyunkyung Kim, and R. J. Birgeneau, Phys. Rev. Lett. {\bf 92},
217004 (2004)

\bibitem{waki_full} S. Wakimoto, R. J. Birgeneau, A. Kagedan, 
Hyunkyung Kim, I. Swainson, K. Yamada, and H. Zhang,
Phys. Rev. B {\bf 72}, 064521 (2005).

\bibitem{Hill_05} J. P. Hill, D. S. Coburn, Y.-J. Kim, T. Gog, N.
Kodituwakku, and H. Sinn, to appear in J. Phys. Chem. Solids

\bibitem{Tsutsui_03} K. Tsutsui, T. Tohyama, and S. Maekawa,
Phys. Rev. Lett. {\bf 91}, 117001 (2003).

\bibitem{Tsutsui_04} K. Tsutsui, T. Tohyama, and S. Maekawa,
Physica C {\bf 412-414}, 143 (2004).

\bibitem{Tsutsui_private} K. Tsutsui (private communication).

\bibitem{Ino_97} A. Ino,  T. Mizokawa, A. Fujimori, K. Tamasaku,
H. Eisaki, S. Uchida, T. Kimura, T. Sasagawa, and K. Kishio,
Phys. Rev. Lett. {\bf 79}, 2101 (1997).


\bibitem{Kuiper_98} Pieter Kuiper, J.-H. Guo, Conny S\.{a}the, 
L.-C. Duda, Joseph Nordgren, J. J. M. Pothuizen, F. M. F. de Groot, 
and G. A. Sawatzky,
Phys. Rev. Lett. {\bf 80}, 5204 (1998).

\bibitem{Ghiringhelli_04} G. Ghiringhelli, N. B. Brookes, E. Annese, 
H. Berger, C. Dallera, M. Grioni, L. Perfetti, A. Tagliaferri, and 
L. Braicovich,
Phys. Rev. Lett. {\bf 92}, 117406 (2004).



\end{references}
\end{document}